\documentclass[pre,aps,onecolumn,superscriptaddress,notitlepage,nofootinbib]{revtex4}
\usepackage{amssymb}
\usepackage{epsfig}
\usepackage{amsmath}
\usepackage{subfigure}
\usepackage{graphicx}
\usepackage{textcomp}
\usepackage{url}
\usepackage{float}
\usepackage{color}
\usepackage{cases}
\usepackage[normalem]{ulem}
\usepackage{xcolor,cancel}

\usepackage[titletoc,title]{appendix}

\usepackage[colorlinks=true, urlcolor=blue, anchorcolor=blue, citecolor=blue,filecolor=blue,linkcolor=blue,menucolor=blue
]{hyperref}

\begin{document}
\title{Geometrical optics of large deviations of Brownian motion in inhomogeneous media}

\author{Tal Bar}
\email{Tal.Bar4@mail.huji.ac.il}

\author{Baruch Meerson}
\email{meerson@mail.huji.ac.il}

\affiliation{Racah Institute of Physics, Hebrew University of Jerusalem, Jerusalem
91904, Israel}

\nopagebreak

\begin{abstract}
Geometrical optics provides an instructive insight into Brownian motion, ``pushed" into a large-deviations regime by imposed constraints.  Here we extend geometrical optics of Brownian motion by accounting for diffusion inhomogeneity in space. We consider three simple model problems of Brownian motion on the line or in the plane in situations where the diffusivity of the Brownian particle depends on one spatial coordinate. One of our results describes ``Brownian refraction": an analog of refraction of light passing through a boundary between two media with different refraction indices.
\end{abstract}
\maketitle
\nopagebreak

\section{Introduction}
\label{intro}

Large deviations of Brownian motion have recently attracted much attention from statistical physicists \cite{Grosberg,Ikeda,Schuss,Nechaev,SM2019,M2019,MS2019a,M2019b,MM2020,Agranovetal,Vladimirov,M2020,Valov}.
One setting where large deviations arise naturally  is stochastic searching of a distant target by multiple random searchers, such as receptors searching for  a reaction site on the surface of a living cell \cite{Coombs}, or sperm cells searching for an oocyte \cite{Eisenbach}. In such systems the multiple searchers essentially compete among themselves for reaching the target first and performing a biological function. When the number of searchers is very large (for example, there are about $3\times 10^8$ sperm cells which attempt to reach the oocyte after copulation in humans), the searcher which succeeds in arriving first does it in an unusually short time. As a result, this searcher's trajectory looks quite differently from \emph{typical} trajectories of a single searcher conditioned on reaching the target at \emph{any} time \cite{Grosberg,Ikeda,Schuss,SM2019,M2019,MS2019a,M2019b,MM2020,Agranovetal,M2020}.

For a single Brownian searcher, the probability of an extremely fast arrival at the target is very small; it corresponds to a short-time tail of the complete probability distribution of the arrival times. Short-time tails of this type can be efficiently described theoretically by the optimal fluctuation method which, in the context of Brownian motion, is called ``geometrical optics of Brownian motion" \cite{Grosberg,Ikeda,Schuss,SM2019,M2019,MS2019a,M2019b,MM2020,Agranovetal,M2020}. An important advantage of geometrical optics is that it allows to account, in a simple manner, for additional constraints. Furthermore, apart from providing the distribution tail of interest, geometrical optics predicts the optimal path of the system, that is the most likely trajectory of the Brownian particle which dominates the probability distribution tail. The optimal path gives a valuable insight into the physics of the large deviation in question.

Previous works on geometrical optics of Brownian motion assumed that the Brownian motion, ``pushed" into a large-deviation regime by imposed constraints, occurs in a homogeneous medium (possibly with some impenetrable regions) \cite{Grosberg,Ikeda,Schuss,SM2019,M2019,MS2019a,M2019b,MM2020,Agranovetal,M2020}.  In some applications, however, the homogeneity assumption is unrealistic. For example, the diffusivity of a reactant macromolecule, exploring a living cell  can
vary quite significantly due to a varying physical and chemical composition of different cell regions \cite{Bressloff}. Here we consider large deviations of a Brownian particle with a space-dependent diffusivity, as described by the Langevin equation
\begin{equation}\label{Langevin}
\dot{\mathbf{x}}=\sqrt{2 D[\mathbf{x}(t)]} \,\boldsymbol{\xi}(t)\,.
\end{equation}
Here $\mathbf{x}(t)$ is the displacement of the particle, $D=D(\mathbf{x})$ is the space-dependent diffusivity, and $\boldsymbol{\xi}(t)$ is the Gaussian white noise with unit norm and zero mean. Using the H\"{a}nggi-Klimontovich interpretation of the multiplicative noise in Eq.~(\ref{Langevin}) \cite{Sokolov}, one can write down the diffusion equation
\begin{equation}\label{diffusioneq}
\frac{\partial P(\mathbf{x},t)}{\partial t}= \nabla \cdot \left[D(\mathbf{x}) \nabla P(\mathbf{x},t)\right]
\end{equation}
for the probability $P(\mathbf{x},t)$ of observing the particle at position $\mathbf{x}$ at time $t$.  In one spatial dimension Eqs.~(\ref{Langevin}) and (\ref{diffusioneq}) were studied in Ref. \cite{key-20} in connection to phenomena accompanying anomalous diffusion, and in Ref. \cite{Farago2014} in the context of fluctuation-dissipation relation.  Here we are interested in large deviations in this system. The starting point of our analysis will be
the Wiener's action
\begin{equation}
-\ln P[\mathbf{x}(t)]\simeq S=\int_{0}^{T}\frac{\dot{\mathbf{x}}^{2}(t)}{4D\left[\mathbf{x}(t)\right]}\,dt\,,
\label{Action}
\end{equation}
which gives the probability $\mathcal{P}[\mathbf{x}(t)]$ of a Brownian path $\mathbf{x}\left(t\right)$ up to pre-exponential
factors. A large-deviation regime corresponds to a very large action, $S\gg 1$. It is this regime where the (tail of the)  probability distribution of interest is dominated by a single optimal path $\mathbf{x}(t)$: a deterministic trajectory which minimizes the Wiener's action (\ref{Action}) over all possible trajectories subject to additional, problem-specific  constraints.

This is the essence of the method of geometrical optics of large deviations of Brownian motion
\cite{Grosberg,Ikeda,Schuss,SM2019,M2019,MS2019a,M2019b,MM2020,Agranovetal,M2020}, previously developed for homogeneous media. For  $D=\text{const}$  (and in the absence of any integral constraints) the action~(\ref{Action}) is minimized by a particle motion with constant speed, $|\dot{\mathbf{x}}|=\text{const}$,
along a geodesic: the shortest path between the initial and final position of the particle, subject to additional constraints, if any.  The action along such a path is
\begin{equation}
S=\frac{1}{4D}\int_{0}^{T}\left(\frac{\mathcal{L}}{T}\right)^{2}\,dt=\frac{\mathcal{L}^{2}}{4DT},\label{eq:action_minimial_length}
\end{equation}
where $\mathcal{L}$  is the path's length. For $D=\text{const}$ the problem therefore reduces
to minimizing $\mathcal{L}$ subject to the additional constraints \cite{MS2019a}.

Here we are interested in inhomogeneous diffusion as described by Eqs.~(\ref{Langevin})-(\ref{Action}). To this end we consider three simple model problems
where $D$ depends only on one spatial coordinate $x$. As we will see, the diffusion inhomogeneity can lead to interesting effects. One of them is the ``Brownian refraction", which emphasizes the geometrical nature of large deviations of Brownian motion and provides an additional justification to the term ``geometrical optics of diffusion".

Here is a layout of the remainder of the paper. In Sec.~\ref{1dpiecewise} we use the geometrical optics to determine the short-time asymptotic of the propagator of the  Brownian motion on the line with a piecewise constant diffusivity $D(x)$. In this case the propagator can be also found exactly by solving Eq.~(\ref{diffusioneq}) (see the Appendix). This allows us to verify geometrical optics' predictions. Importantly, the latter also include the
optimal path of the system, which is not directly available from the exact solution. In Sec. \ref{2dpiecewise} we find the short-time asymptotic of the propagator of the Brownian motion on the plane with a piecewise constant $D(x)$. It is here where we encounter the Brownian refraction. In Sec. \ref{continuousD} the geometrical optics is used to determine the short-time asymptotic of the propagator of the  Brownian motion on the line with a continuously varying diffusivity $D(x)$. A brief discussion of our results and of their possible extensions is included in Sec. \ref{Discussion}.

\section{Brownian motion on the line with a piecewise constant $D(x)$}
\label{1dpiecewise}

Our first example is Brownian motion on a line with a piecewise constant
diffusivity:
\begin{equation}
D\left(x\right)=\begin{cases}
D_{1}, & x<\ell\,, \\
D_{2}\equiv\kappa^{2}D_{1}, & x>\ell,
\end{cases}\label{eq:1D diffusion factor}
\end{equation}
where $\ell>0$. The particle starts at the origin at $t=0$, and we are interested in the short-time, $T\to 0$, asymptotic of the probability density $P(L,T)$ of observing the particle at a point $x=L>\ell$ at time $T>0$.

On each of the intervals $0<x<\ell$ and $\ell<x<L$ the diffusivity is constant. Therefore, in view
of the above-mentioned geometrical optics results for $D=\text{const}$,
the optimal path $x(t)$ must be a piecewise linear function of time. Let us denote by $\tau$
the (\textit{a priori} unknown) time of crossing the boundary $x=\ell$ between the
two constant-$D$ regions (we assume that this time is unique.)  Demanding that $x(t)$ obey
the conditions $x(0)=0$ and $x(T)=L$, we obtain
\begin{equation}
x\left(t\right)=\begin{cases}
\frac{\ell t}{\tau}, & 0<t<\tau\,, \\
\frac{\ell (T-t)+L (t-\tau )}{T-\tau }, & \tau<t<T\,.
\end{cases}\label{piecewiseansatz}
\end{equation}
The action on this composite trajectory is the sum of contributions, described by Eq.~(\ref{eq:action_minimial_length}), from each of the two intervals:
\begin{equation}\label{general tau action}
S \simeq  \frac{\ell^{2}}{4D_{1}\tau}+\frac{\left(L-\ell\right)^{2}}{4D_{2}\left(T-\tau\right)}
  = \frac{1}{4}\left(\frac{t_{1}}{\tau}+\frac{t_{2}}{T-\tau}\right)\,,
\end{equation}
where
\begin{equation}\label{times1}
t_{1}=\frac{\ell^{2}}{D_{1}}\quad \mbox{and}\quad t_{2}=\frac{\left(L-\ell\right)^{2}}{D_{2}}=\frac{\left(L-\ell\right)^{2}}{\kappa\sqrt{D_{1}}}
\end{equation}
are the characteristic diffusion times through each of the two media. The minimum action is achieved for
\begin{equation}\label{tau1min}
\tau=\frac{\sqrt{t_{1}}}{\sqrt{t_{1}}+\sqrt{t_{2}}}\,T=\frac{T}{\frac{1}{\kappa}\left(\frac{L}{\ell}-1\right)+1}\,,
\end{equation}
and we obtain the short-time asymptotic
\begin{equation}
 \label{approximate soiution}
  -\ln P(L,T\to 0)\simeq  S_{\text{min}}
  =\frac{\left(\sqrt{t_{1}}+\sqrt{t_{2}}\right)^{2}}{4T}
  = \frac{\left[L-\left(1-\kappa\right)\ell\right]^{2}}{4D_{2}T}\,.
\end{equation}
The same $T\to 0$ asymptotic (\ref{approximate soiution}) follows from exact solution of this problem (see the Appendix for details):
\begin{align}
P_{\text{exact}}\left(L,T\right) & =\frac{1}{\sqrt{\pi T}\left(\sqrt{D_{1}}+\sqrt{D_{2}}\right)}e^{-\frac{\left[L-\left(1-\kappa\right)\ell\right]^{2}}{4D_{2}T}}\,.\label{exact solution}
\end{align}
Indeed, when $T\to 0$, the pre-exponential factor in Eq.~(\ref{exact solution}) (which is missed by the leading-order geometrical optics calculations) becomes subleading, and Eqs.(\ref{approximate soiution}) and (\ref{exact solution}) coincide in the leading order.

\begin{figure}[ht]
\includegraphics[scale=0.40]{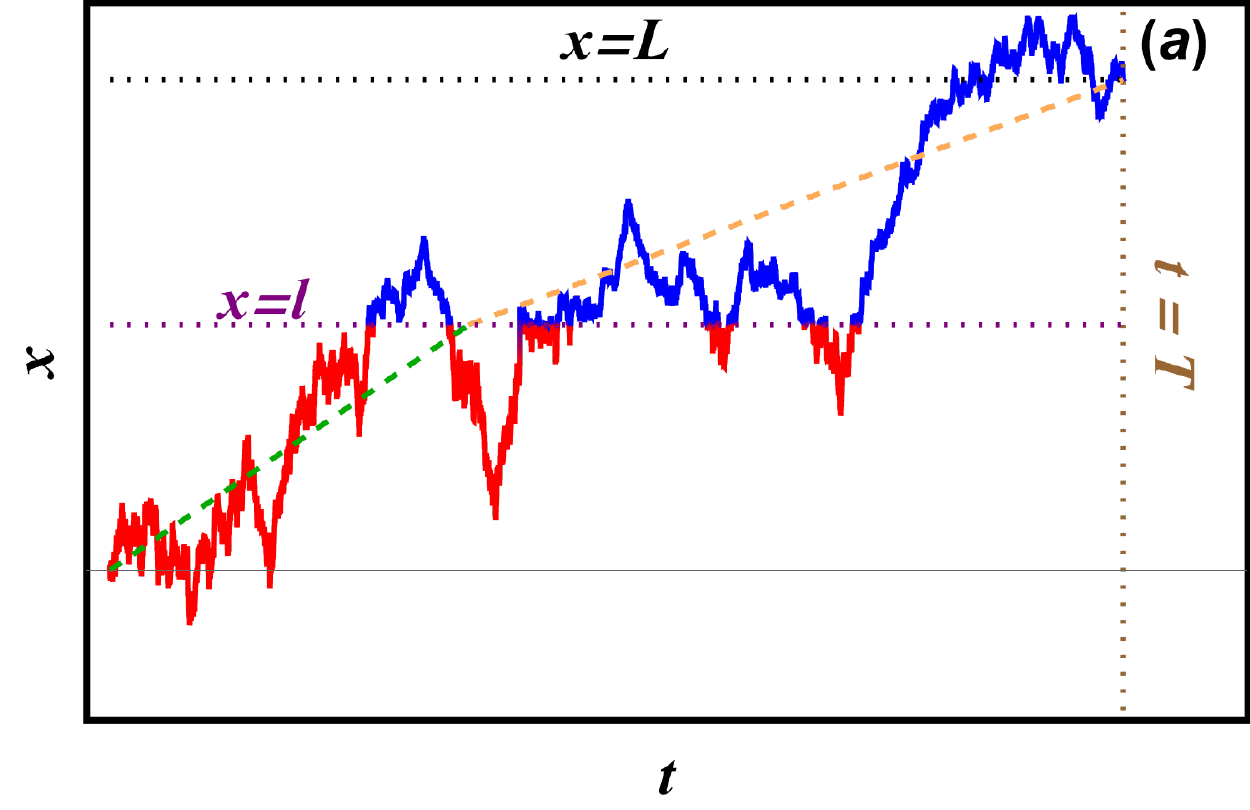}
\includegraphics[scale=0.40]{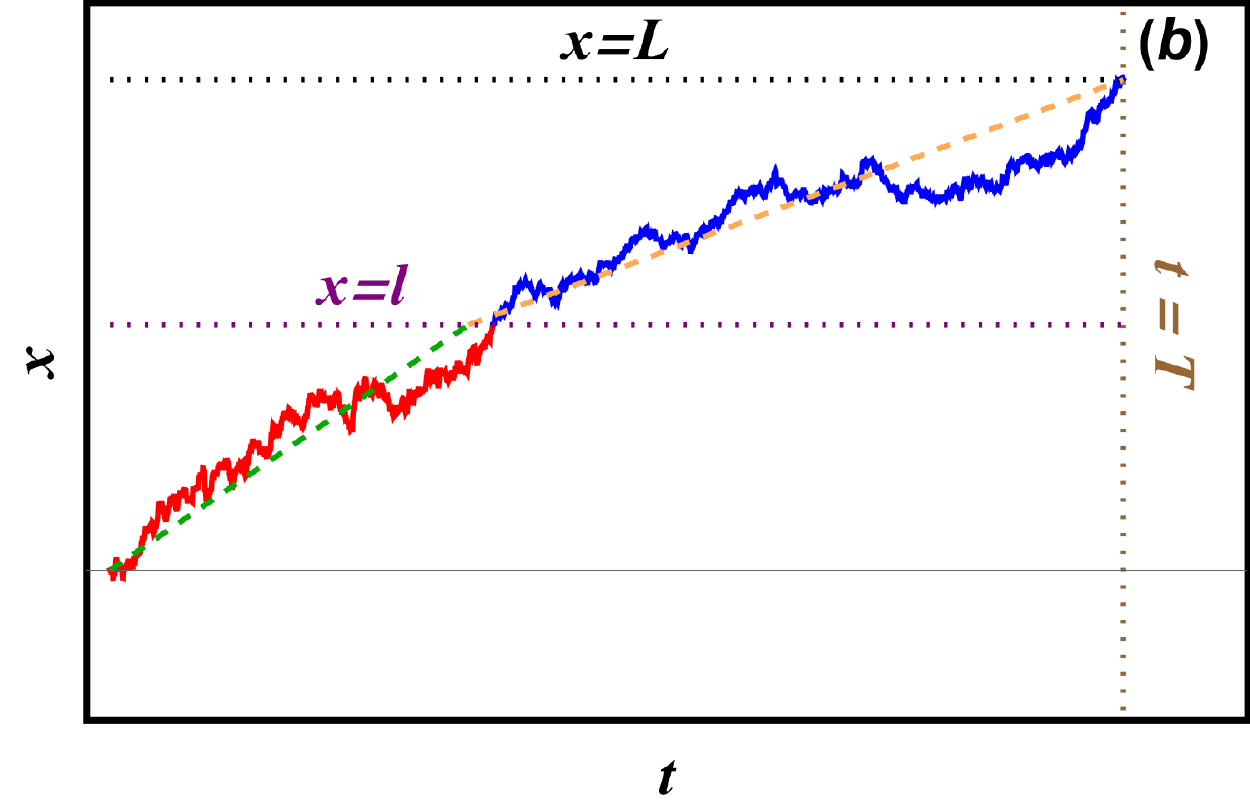}
\caption{Solid lines: two simulated Brownian paths for the setting of Sec. \ref{1dpiecewise}. The parameters are
$L=6$, $\ell=3$, $D_1=10$, $D_2=3$,
and $T=6 $ (a) and $T=0.45$ (b).
The dashed lines show the piecewise linear optimal path predicted by geometrical optics, Eqs.~(\ref{piecewiseansatz}) and (\ref{tau1min}).}
\label{1dfig}
\end{figure}

Figure \ref{1dfig} a and b  gives two examples of simulated Brownian paths in this setting. The particle starts at the origin and arrives at $x=L$ at two different times $T$. The figure also shows  the piecewise linear optimal path $x(t)$ at $T\to 0$, predicted by Eqs.~(\ref{piecewiseansatz}) and (\ref{tau1min}) for the same $D_1$ and $D_2$. As $T$ decreases, the simulated paths start approaching the predicted short-time optimal path.

Notice that the  geometrical optics result (\ref{approximate soiution}) gives a correct leading-order asymptotic $P(L,T\to 0)$ for a whole class of settings dealing with a Brownian particle starting at the origin and arriving at $x=L$ at $t=$. One such setting -- when the Brownian motion is defined on the whole line -- is solved exactly in the Appendix. But one could actually consider
a Brownian motion on any interval $(a,b)$ which includes in itself the interval $0<x<L$, with either reflecting or absorbing boundary conditions at the boundaries $x=a$ and $x=b$. In this case the exact solution would of course be different from Eq.~(\ref{exact solution}), but its leading-order $T\to 0$ asymptotic will still be described by Eq.~(\ref{approximate soiution}). This is explained by the character of the optimal path: when conditioned on reaching $x=L$ in a very short time,
the particle cannot spend any time on exploring regions beyond the interval $(0,L)$, rendering them irrelevant for the leading-order calculations.

\section{Brownian motion on the plane with a piecewise constant $D(x)$: Brownian refraction}
\label{2dpiecewise}

\begin{figure}[ht]
\includegraphics[scale=0.40]{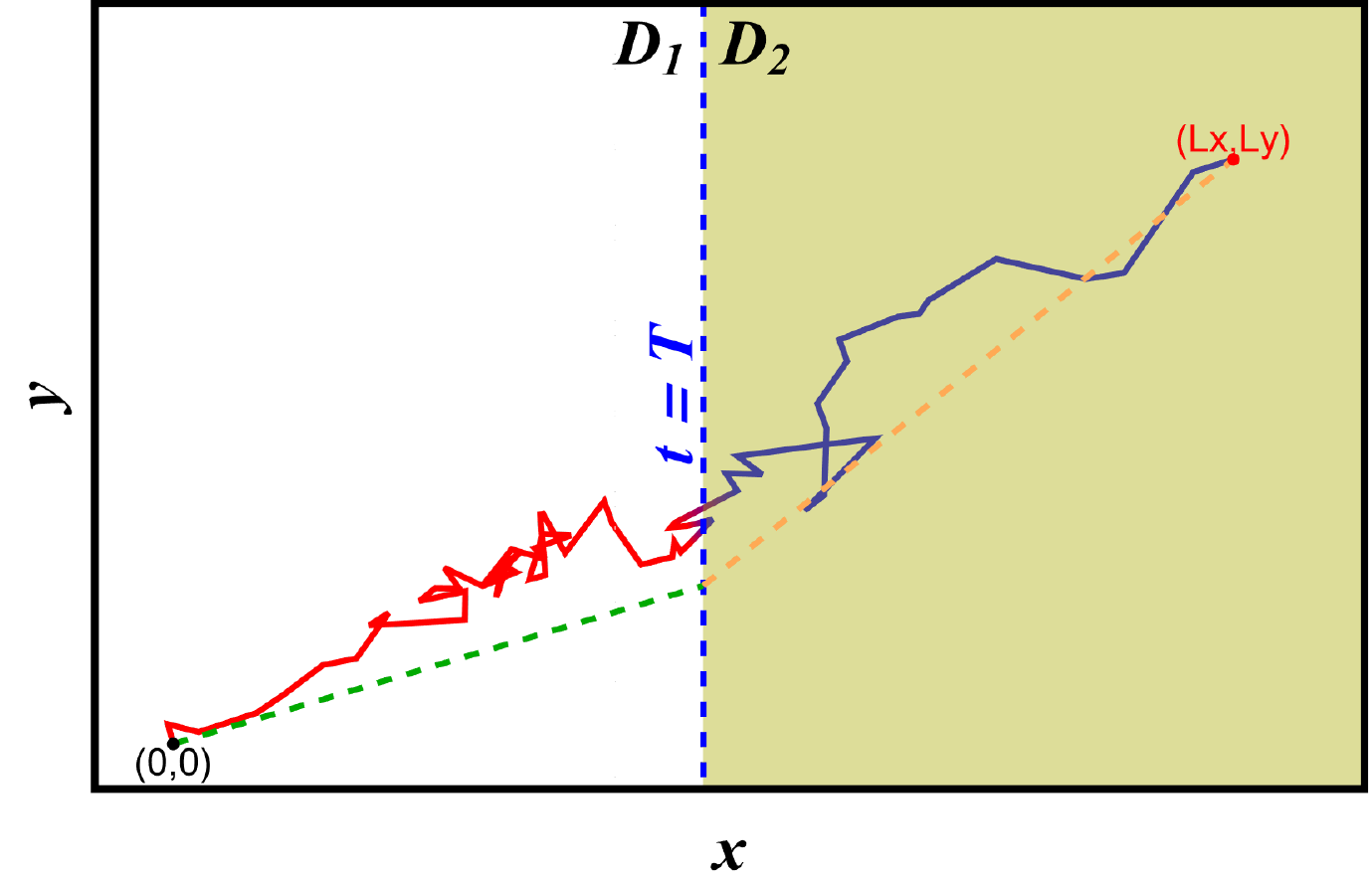}
\caption{A simulated Brownian path for the setting of Sec. \ref{2dpiecewise} with $\kappa>1$. The parameters are $L_x=L_y=10$, $\ell=5$, $D_1=1$, $D_2=3$, and  $T =4.17$ (solid line). The dashed lines show the piecewise linear
optimal path predicted by the geometrical optics in the limit of $T\to 0$. This path obeys the Snell's law (\ref{eq:Snell's law}).}
\label{fig:2d snell}
\end{figure}

Now let us consider a Brownian motion in the $(x,y)$ plane with a  piecewise constant diffusivity
\begin{equation}
D\left(x\right)=\begin{cases}
D_{1}\,, & x<\ell\,,\\
D_{2}\equiv\kappa^{2}D_{1}\,, & x>\ell\,,
\end{cases}\label{eq:2D diffusion factor-1}
\end{equation}
for any $y$. The Brownian particle starts
at the origin at $t=0$, and
we are interested in the short-time asymptotic, $T\to 0$, of the probability $P(L_x,L_y,T)$ for the particle to reach the point
$(L_x,L_y)$, so that $L_{x}>\ell$, at time $t=T$, see Fig. \ref{fig:2d snell}.

We assume that the particle crosses the boundary $x=\ell$ between the two regions with different diffusivities at a unique time
$0<\tau<T$, and the $y$-coordinate of the particle at this moment is $y_{0}$. The action on this trajectory is the sum of two terms, described by Eq.~(\ref{eq:action_minimial_length}), which come from each of the two media:
\begin{equation}
 S(\tau,y_0)=\frac{1}{4}\left[\frac{\ell^{2}+y_{0}^{2}}{D_{1}\tau}+
 \frac{\left(L_{x}-\ell\right)^{2}+\left(L_{y}-y_{0}\right)^{2}}{D_{2}\left(T-\tau\right)}\right]
  =\frac{1}{4}\left(\frac{t_{1}}{\tau}+\frac{t_{2}}{T-\tau}\right)\,, \label{general tau action-1}
\end{equation}
with obvious definitions of the characteristic diffusion times $t_1$ and $t_2$ for each of the two media.
Now we should minimize the action (\ref{general tau action-1}) with respect to the two auxiliary parameters $\tau$ and $y_0$. It is convenient to do it sequentially. Notice that Eqs.~(\ref{general tau action-1}) and (\ref{general tau action}) have
the same form in terms of $\tau$, $t_{1}$ and $t_{2}$. Therefore, at fixed $y_0$, the minimum of the action with respect to $\tau$
is achieved at $\tau=\frac{\sqrt{t_{1}}}{\sqrt{t_{1}}+\sqrt{t_{2}}}T$ [see Eq.~(\ref{tau1min})], and we obtain
\begin{equation}
  S(y_0)= \frac{1}{4T}\left(\sqrt{t_{1}}\!+\!\sqrt{t_{2}}\right)^{2}
   \!=\!\frac{1}{4T}\!\left[\sqrt{\frac{\ell^{2}\!+\!y_{0}^{2}}{D_{1}}}
   +\sqrt{\frac{\left(L_{x}\!-\!\ell\right)^{2}\!+\!\left(L_{y}\!-\!y_{0}\right)^{2}}{D_{2}}}\right]^{2}.
   \label{eq:genral y0 solution}
\end{equation}
The minimization with respect to the remaining parameter $y_0$ demands $dS(y_0)/dy=0$, which leads to the relation
\begin{equation}
\frac{L_{y}-y_{0}}{\sqrt{D_{2}}\sqrt{\left(L_{x}-l\right)^{2}+\left(L_{y}
-y_{0}\right)^{2}}}=\frac{y_{0}}{\sqrt{D_{1}}\sqrt{\ell^{2}+y_{0}^{2}}}\label{eq:constraint}
\end{equation}
To better understand this result we notice that $y_{0}$  and  $L_{y}-y_{0}$
can be written as $y_{0}=\ell\tan\theta_{1}$, and $L_{y}-y_{0}=\left(L_{x}-\ell\right)\tan\theta_{2}$, respectively, where $\theta_{1}$ ($\theta_{2}$) is the angle between
the particle trajectory in the left (right) region and the straight line $y=y_{0}$. In other words,
$\theta_{1}$ and $\theta_2$ are the angles of incidence and refraction, respectively. In the new notation
Eq.~(\ref{eq:constraint})
becomes, remarkably,
\begin{equation}
\frac{\sin \theta_1}{\sin \theta_2}= \frac{\sqrt{D_1}}{\sqrt{D_2}} =\frac{1}{\kappa}\,,
\label{eq:Snell's law}
\end{equation}
providing an unexpected Brownian-motion analog of Snell's law of optics (the latter
describes the refraction of light or other waves passing through a boundary between two media with different refraction indices \cite{Crawford,Feynman}).
Notice that the role of the refraction index $n$ in optics is played here by $1/\sqrt{D}$.
The ``Brownian refraction", described by Eq.~(\ref{eq:Snell's law}), provides a further justification to the term
``geometrical optics of
(large deviations of) Brownian motion" \cite{Grosberg,Ikeda,Schuss,SM2019,M2019,MS2019a,M2019b,MM2020,Agranovetal,M2020}\footnote{\label{different}To remind the reader, the Snell's law of refraction can be derived by minimizing the arrival time of the ray to the target point \cite{Feynman}. For the Brownian motion the arrival time is fixed, and one minimizes the Wiener's action (\ref{Action}). In the light of these differences the Brownian refraction phenomenon is somewhat surprising.}.

Let us complete the  solution. The relation~(\ref{eq:constraint}) yields a
quartic equation for $y_0$, the analytical solutions of which is too bulky
to work with. It is more convenient to present the solution in a parametric form.
Using Eq.~(\ref{eq:constraint}), we can express $L_{y}$ via $y_0$:
\begin{equation}
L_{y}=y_{0}\left[1+\frac{\left(L_{x}-l\right)\kappa}{\sqrt{l^{2}+\left(1-\kappa^{2}\right)y_{0}^{2}}}\right].
\label{eq:to ly solutions}
\end{equation}
Plugging this expression into Eq. (\ref{eq:genral y0 solution}), we obtain
\begin{equation}
S=\frac{\ell^{2}+y_{0}^{2}}{4D_{2}T}\left[\kappa+\frac{\left(L_{x}-\ell\right)}{\sqrt{\ell^{2}
+\left(1-\kappa^{2}\right)y_{0}^{2}}}\right]^{2}\,. \label{subtituted eqn}
\end{equation}
Equations~(\ref{eq:to ly solutions}) and~(\ref{subtituted eqn}) yield $-\ln P \simeq S$ (which we are after) in a parametric form, where $y_0$ serves as the parameter.

It is instructive to look at the limits of $S$ at $\kappa\to 0$ and $\kappa \to \infty $ at fixed $D_2$:
\begin{equation}
S\left(\kappa\right)\simeq\begin{cases}
\frac{\left(L_{x}-\ell\right)^{2}}{4D_{2}T}\,, & \kappa \to 0\,,\\
\frac{\ell^{2}}{4D_{1}T}\,, & \kappa \to \infty\,.
\end{cases}
\end{equation}
The corresponding optimal paths in these two limits are the following. At $\kappa \to 0$ we have $y_0\to L_y$, and $\tau\to 0$. The form of the trajectory is
\begin{equation}
y(x,\kappa\to 0)\simeq \begin{cases}
\frac{L_y}{\ell}\,x\,, & 0<x<\ell\,,\\
L_y\,, & \ell<x<L_x\,.
\end{cases}
\end{equation}
Here the diffusion in media 1 is fast, and the particle reaches the point $(\ell,L_y)$  almost immediately
so as to minimize the distance it has to travel in media 2, where the diffusivity is finite. The particle spends almost
all of the allocated time $T$ in media 2.

As $\kappa \to \infty$, $y_0$ goes to zero, and $\tau$ approaches $T$. Now the form of the trajectory is
\begin{equation}
y(x,\kappa\to \infty)\simeq \begin{cases}
0\,, & 0<x<\ell\,,\\
\frac{L_y(x-\ell)}{L_x-\ell}\,, & \ell<x<L_x\,.
\end{cases}
\end{equation}
Here the diffusion in media 1 is slow. Therefore,  the particle moves along the $x$-axis so as to minimize the distance it has to travel
here, and its spends almost all of the allocated time $T$ to arrive at $x=\ell$. Then it almost immediately ``jumps" to the point
$(L_x,L_y)$ along the straight line.

\section{Brownian motion on the line with a continuously varying $D(x)$}
\label{continuousD}

Here we return to one spatial dimension and suppose that a Brownian particle is released at $t=0$ at $x=0$ in a medium with a
continuously varying diffusivity $D\left(x\right)$.

We are interested in the short-time asymptotic of the
probability $P(L,T)$ of observing the particle at $x=L$ at time $t=T$. Using geometrical optics, we
identify the Lagrangian
\begin{equation}\label{Lagrangian}
\frak{L}\left(x,\dot{x}\right) = \frac{\dot{x}^2}{4 D(x)}
\end{equation}
of the Wiener's action (\ref{Action}). Since this Lagrangian does not depend explicitly on time, there is an integral of motion which, in this case, coincides with the Lagrangian itself, $\frak{L}\left(x,\dot{x}\right) =\text{const}$. Therefore, we can write
\begin{equation}
\dot{x}=C \sqrt{\ D\left(x\right)}\,,\label{gen d solution}
\end{equation}
where $C$ is a constant to be determined. A comparison of Eq.~(\ref{gen d solution}) with the Langevin equation~(\ref{Langevin})
shows that  the optimal realization of the rescaled Gaussian white noise $\xi(t)$, conditioned on a very fast arrival of the Brownian particle at $x=L$, is simply a constant, equal to $C/\sqrt{2}$.

Equation~(\ref{gen d solution}) can be solved in quadratures for any $D(x)$, and we obtain
\begin{equation}\label{quadrature}
C t = \int_0^x \frac{dy}{\sqrt{D(y)}}\,,
\end{equation}
where we used the boundary condition $x(0)=0$. Using the second boundary condition $x(T)=L$, we determine the constant $C$:
\begin{equation}\label{findC}
C=\frac{1}{T}\int_0^L\frac{dx}{\sqrt{D(x)}}\,.
\end{equation}
Equations~(\ref{quadrature}) and (\ref{findC}) yield the optimal path $x(t)$ in an implicit form for any $D(x)$.
Further, plugging Eq.~(\ref{gen d solution}) into Eq.~(\ref{Action}) and using Eq.~(\ref{findC}), we obtain a  general expression for the action:
\begin{equation}
S=\int_{0}^{T}\frac{\dot{x}^{2}}{4D\left(x\right)}\,dt=\frac{1}{4T}\left[\int_0^L\frac{dx}{\sqrt{D(x)}}\right]^2\,.\label{eq:simple action}
\end{equation}

\begin{figure}[ht]
\includegraphics[scale=0.35]{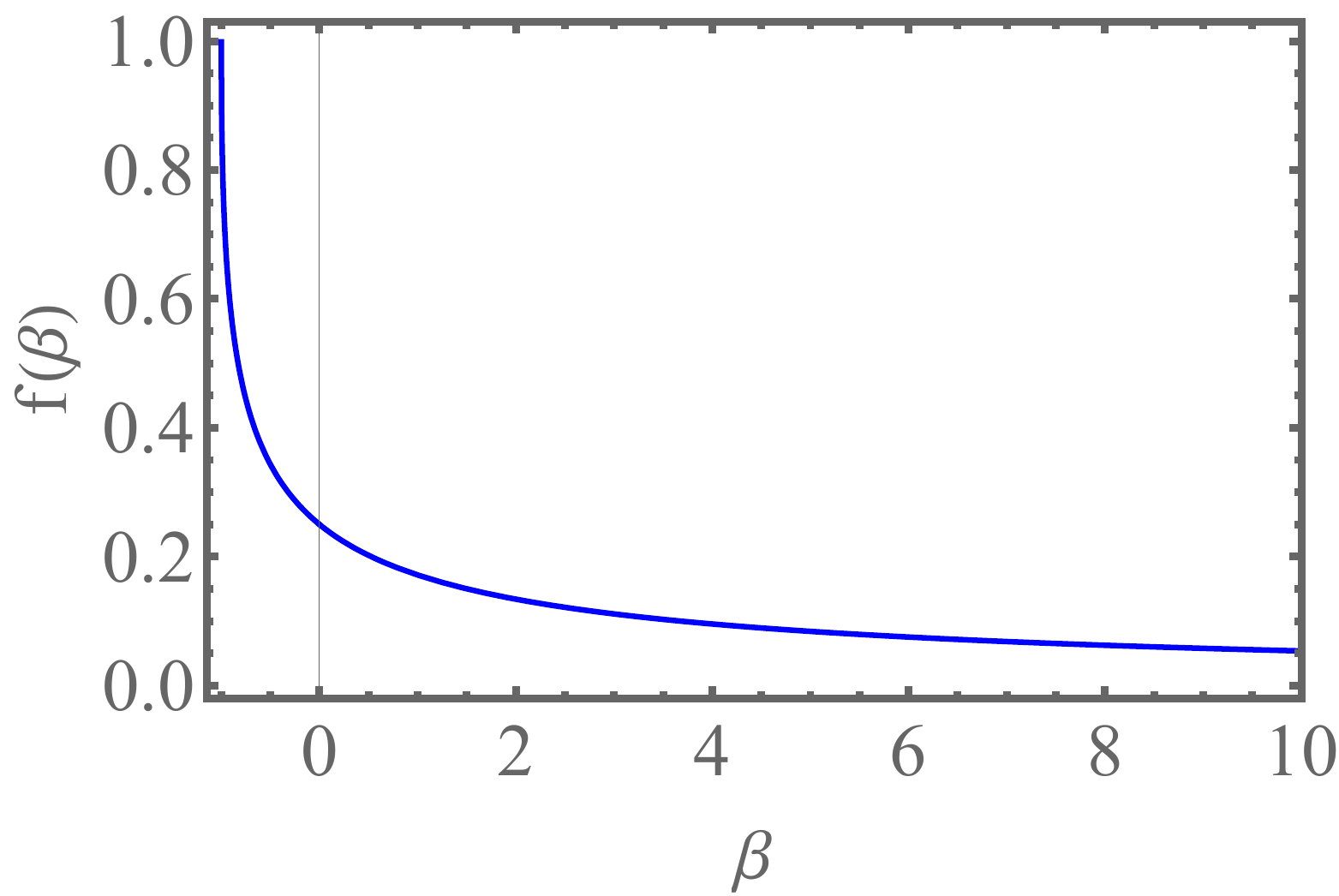}
\caption{The function $f(\beta)$ from Eq.~(\ref{eq:smin}).}
\label{fvsbeta}
\end{figure}

As a simple example, consider a linear diffusivity profile, $D(x) =D_0 (1+\beta x/L)$, where we must assume $\beta>-1$ in order to keep $D(x)$ positive in the relevant region $0<x<L$. Using Eqs.~(\ref{quadrature}), 
we obtain
\begin{equation}
\label{x(t)linear}
x(t)=L \frac{t}{T} \left[\frac{t}{T}+\frac{2
   \left(\sqrt{\beta +1}-1\right)}{\beta}\left(1-\frac{t}{T}\right)
   \right]\,.
\end{equation}
The action~(\ref{eq:simple action}) takes the form
\begin{equation}
S=\frac{L^2}{D_0 T}\,f(\beta)\,,\;\mbox{where}\;\: f(\beta)=  \frac{\left(\sqrt{\beta +1}-1\right)^2}{\beta ^2}\,. \label{eq:smin}
\end{equation}
It gives the short-time asymptotic of $P(L,T)$ via the relation $-\ln P(L,T) \simeq S$.
A graph of the function $f(\beta)$ is shown in Fig.~\ref{fvsbeta}. As one can see, the action decreases monotonically with an increase of $\beta$. It reaches its maximum value $L^2/(DT)$  at $\beta=-1$. In this case the diffusivity vanishes at $x=L$. As $\beta$ goes to infinity, $f(\beta)$ goes down as $1/\beta$. Finally, at $\beta\to 0$  we reproduce the constant-diffusivity action $S=L^2/(4 D_0 T)$.

\begin{figure}[ht]
\includegraphics[scale=0.35]{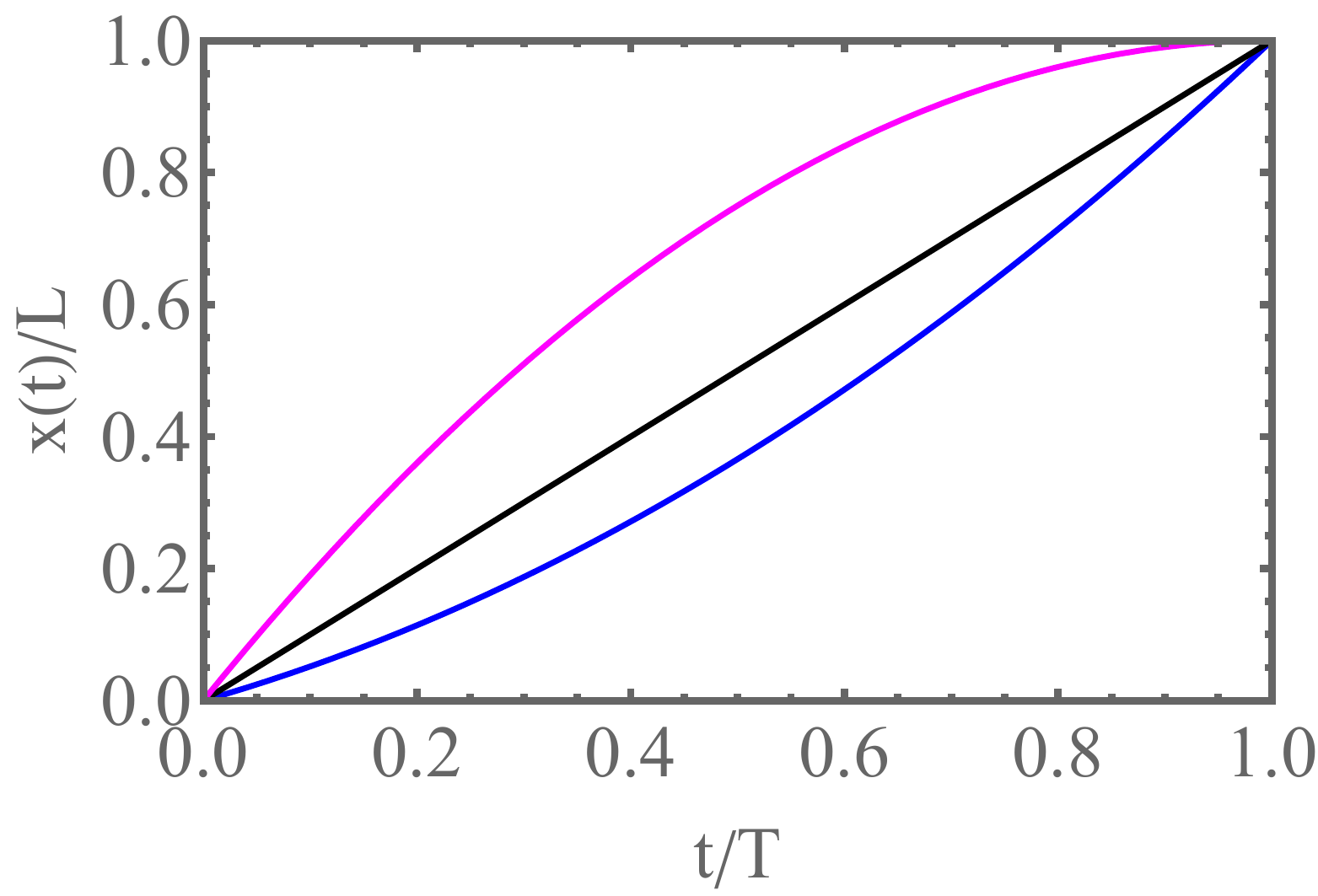}
\caption{The optimal path $x(t)$, as described by Eq.~(\ref{x(t)linear}), for the linear diffusivity profile $D(x)=D_0 (1+\beta x/L)$ with $\beta=10$ (blue) and $-1$ (magenta). The black line shows $x(t)$ for the constant diffusivity, $\beta=0$.}
\label{x(t)linearD}
\end{figure}

The optimal path $x(t)$, as described by Eq.~(\ref{x(t)linear}), is shown in Fig.~\ref{x(t)linearD}  for three different values of $\beta$. In general, $x(t)$ is a parabola: concave or convex
depending on whether $\beta$ is positive or negative, respectively.  For $\beta=0$ (a constant diffusivity) $x(t)$ is a straight line as to be expected.

\section{Discussion}
\label{Discussion}

We considered three simple model problems, illustrating the role of spatial variations of the diffusivity in large deviations of Brownian motion.  One of our predictions is what can be called ``Brownian refraction": a close analog of refraction of light passing through a boundary between two media with different refraction indices. This phenomenon appears when one conditions the Brownian particle on arrival at a distant point in a very short time.

It would be interesting to extend the geometrical-optics approach by accommodating additional local or integral constraints, or constraints in the form of inequalities, like it was done for the homogeneous diffusion \cite{Grosberg,Ikeda,Schuss,SM2019,M2019,MS2019a,M2019b,MM2020,Agranovetal,M2020}

When dealing, in Sec. \ref{2dpiecewise}, with  a Brownian motion on the plane, we assumed that the diffusivity $D$ depends only on one spatial coordinate $x$, and this dependence is piecewise constant. This theory can be extended to a continuously varying $D(x)$. Indeed, the Lagrangian of Eq.~(\ref{Action}),
\begin{equation}\label{2dL1}
\frak{L}(x,\dot{x},\dot{y}) = \frac{\dot{x}^2}{4 D(x)}+\frac{\dot{y}^2}{4 D(x)}
\end{equation}
describes an effective classical mechanics with two degrees of freedom. This system possesses two integrals of motion. The first of them is the Lagrangian itself,
\begin{equation}\label{energyint}
\frac{\dot{x}^2}{4 D(x)}+\frac{\dot{y}^2}{4 D(x)} = E =\text{const}\,,
\end{equation}
An additional integral of motion, $\dot{y}/D[x(t)] = \text{const}$, follows from the fact that the Lagrangian
(\ref{2dL1}) does not depend on $y$. The presence of two integrals of motion allows one to solve the problem in quadratures for any $D(x)$.

Extensions of the theory to a more general case, where
$D(x,y)$ depends both on $x$, and on $y$, encounter an immediate difficulty \cite{coordinates}, because in this case the geometrical-optics description of the system involves only one integral of motion: the Lagrangian $\frak{L}(x,y,\dot{x},\dot{y})$ itself. The lack of additional independent integral of motion makes the geometrical-optics formulation non-integrable analytically \cite{Tabor}. Numerical solutions, however, are certainly possible. In addition, perturbative analytical treatments can be possible if there is a small parameter in the coordinate dependence $D(x,y)$. It would be interesting to explore these possibilities in a future work.

Finally, geometrical optics of large deviations of Brownian motions present only one, although important, example of a class of systems which can be studied with the optimal fluctuation method (OFM). In recent years the OFM has been employed for analysis of large deviations in additional Markovian \cite{Touchette} and non-Markovian \cite{M2019,M2022,MO2022,M2023} stochastic processes. Another well-known large-deviation technique, based on the Feynman-Kac formula \cite{DV,EG}, applies to a certain class of large deviations of \emph{time-averaged} quantities. There is also a group of large-deviation methods, in different areas of physics, which rely on the ``single-big-jump" principle, see Ref. \cite{bigjump} and references therein. A skillful use of all these methods (and sometimes of their combinations \cite{Smith}) should help develop a better understanding of many important, fascinating and,  quite often, non-intuitive large deviations and rare events.

\section*{Acknowledgment}

We thank Oded Farago for useful comments. BM was supported by the Israel Science Foundation (Grant No. 1499/20).

\vspace{0.5 cm}

\appendix
\section{Exact solution for Sec. \ref{1dpiecewise}}

Here we present exact solution of the equation
\begin{equation}\label{diffusioneq2}
\frac{\partial P(x,t)}{\partial t}= \frac{\partial}{\partial x}\left[D\left(x\right)\frac{\partial P}{\partial x}\right]
\end{equation}
for the probability density $P(x,t)$ on an infinite line $|x|<\infty$, subject to the initial condition
\begin{equation}\label{incond}
P\left(x,t=0\right)  =\delta\left(x\right)\,.
\end{equation}
The piecewise constant diffusivity $D(x)$ is given by
\begin{equation}
D\left(x\right)=\begin{cases}
D_{1}\,, & x<\ell\,,\\
D_{2}\,, & x>\ell\,,
\end{cases}
\end{equation}
so we can solve the problem separately in the two regions,
\begin{equation}
\frac{\partial P}{\partial t}=\begin{cases}
D_{1}\frac{\partial^{2}P}{\partial x^{2}}\,, &-\infty< x<\ell\,,\\
D_{2}\frac{\partial^{2}P}{\partial x^{2}}\,, & \ell<x<\infty\,,
\end{cases}
\end{equation}
and properly match the solutions at $x=\ell$. Let us introduce the functions $f(x,t)$ and $g(x,t)$ as follows:
\begin{equation}
P\left(x,t\right)=\begin{cases}
f\left(x,t\right)\,, & x<\ell\,,\\
g\left(x,t\right)\,, & x>\ell\,.
\end{cases}
\end{equation}
The continuity of the probability density $P(x,t)$ and of the probability flux $-D(x)\frac{\partial P(x,t)}{\partial x}$
at $x=\ell$ yield two conditions:
\begin{equation}\label{matching}
f\left(\ell,t\right)  =g\left(\ell,t\right)\,,\quad
D_{1}\frac{\partial f}{\partial x}\Big|_{x=l}  =D_{2}\frac{\partial g}{\partial x}\Big|_{x=l}\,.
\end{equation}
The problem can be solved by Laplace transform. Alternatively, one can look for the solutions in the two regions as
sums of two rescaled and shifted Gaussians. As one can check explicitly, the resulting solution (which is not new, see \textit{e.g.} Ref.~\cite{Farago2020}) is the following:
\begin{equation}
P\left(x,t\right)\!=\!\!\begin{cases}
\frac{1}{\sqrt{4\pi D_{1}t}}\left[\!e^{-\frac{x^{2}}{4D_{1}t}}\!+\!\frac{\sqrt{D_{1}}
-\sqrt{D_{2}}}{\sqrt{D_{1}}+\sqrt{D_{2}}}e^{-\frac{\left(x-2l\right)^{2}}{4D_{1}t}}\right]\!,&
\!x<\ell,\\
\frac{1}{\sqrt{\pi t}\left(\sqrt{D_{1}}+\sqrt{D_{2}}\right)}e^{-\frac{\left[x-\ell\left(1-\sqrt{\frac{D_{2}}{D_{1}}}\right)\right]^{2}}{4D_{2}t}}\!, & \!x>\ell.\label{exactsol}
\end{cases}
\end{equation}
In Sec.~\ref{1dpiecewise} we compare the  exact result (\ref{exactsol}) for $x>\ell$ with our prediction from geometrical optics, see Eq.~(\ref{exact solution}).


\begin{thebibliography}{100}
\bibitem{coordinates} Integrability is preserved if one can perform a coordinate transformation so that $D$ depends only on one of the new coordinates.
\end{thebibliography}
\end{document}